\documentstyle[11pt,paspconf,psfig]{article}

\begin{document}
\newcommand\Msun {M_{\odot}\ } \def\ltsima{$\; \buildrel < \over \sim
\;$} \def\simlt{\lower.5ex\hbox{\ltsima}} \def\gtsima{$\; \buildrel >
\over \sim \;$} \def\simgt{\lower.5ex\hbox{\gtsima}}

\title{Detailed Star-Formation Histories of Nearby Dwarf Irregular
Galaxies using HST\altaffilmark{1}}

\author{Eline Tolstoy}
\affil{European Southern Observatory, Garching, Germany}

\altaffiltext{1}{
Invited Review, at IAU Symposium 192,
``The Stellar Content of Local Group Galaxies'', Sept 1998,
Cape Town, South Africa, eds. Whitelock \& Cannon, in press
}

\begin{abstract}
Unless the nearby universe is subtly anomalous, it should contain a 
relatively normal selection of galaxies whose histories are representative 
of field galaxies in general throughout the Universe.  We can therefore 
take advantage of our ability to resolve the nearby galaxies into 
individual stars to directly, and accurately, measure star formation 
histories.  The star formation histories are determined from model-fitting 
analysis, based on stellar evolution tracks, of colour-magnitude diagrams.  
The most accurate information on star formation rates extending back to the 
earliest epoches can be obtained from the structure of the main sequence.  
However, the oldest main sequence turnoffs are very faint, and it is often 
necessary to use the more evolved stars to infer the star formation 
history.  These can then be compared with the properties of galaxies seen 
over a large range of lookback times at moderate to high redshifts.  There 
is considerable evidence that the faint blue galaxies seen in large 
numbers in cosmological redshift surveys are the progenitors of the small
late-type irregular galaxies seen in copious numbers in the Local Group,
and we concentrate on these here.

\end{abstract}



\vskip-1.5cm
\section{Introduction}
The study of resolved stellar populations provides a powerful tool to 
follow galaxy evolution directly in terms of physical parameters such as 
age (star formation history, SFH), chemical composition and enrichment 
history, initial mass function, environment, and dynamical history of the 
system.  Photometry of individual stars in at least two filters and the 
interpretation of Colour-Magnitude Diagram (CMD) morphology gives the least 
ambiguous and most accurate information about variations in star formation 
rate ({\it sfr}) within a galaxy back to the oldest stars (see Figure~1).  
Some of the physical parameters that affect a CMD are strongly correlated, 
such as metallicity and age, since successive generations of star formation 
may be progressively enriched in the heavier elements.  Careful, detailed 
CMD analysis is a proven, uniquely powerful approach ({\it e.g.}, Tosi {\it et 
al.} 1991; Tolstoy \& Saha 1996; Aparicio {\it et al.} 1996; Mighell 1997; 
Dohm-Palmer {\it et al.} 1997a,b; Hurley-Keller {\it et al.} 1998; 
Gallagher {\it et al.} 1998; Tolstoy {\it et al.} 1998; Tolstoy 1998) that 
benefits enormously from the high spatial resolution, high quality imaging.  
$HST$ has led the way in recent years, and we are optimistic that ground 
based telescopes capable of achieving excellent seeing, such as VLT, will 
also provide much useful data, especially in the blue, and in relatively 
sparse stellar systems such as dwarf irregular (dI) galaxies.

The small dI and dSph galaxies are considered the most likely connection to 
higher redshift, late type evolving systems, which makes them the most 
interesting to look at when trying to make a connection to the high 
redshift universe.  In the LG they appear to exhibit a wide variety of 
SFHs.  These results have affected our understanding of galaxy formation 
and evolution by demonstrating the importance of episodic star formation in 
nearby low mass galaxies.  The larger galaxies in the LG have evidence of 
sizeable old halos, which appear to represent the majority of star 
formation in the LG by mass, although the problems distinguishing between 
effects of age and metallicity in a CMD result in a degree of uncertainty 
in the exact age distribution in these halos.  High quality, deep imaging 
is the only way of extending detailed SFH studies past the Galaxy and its 
satellites, enabling us to obtain a picture of the fossil record of star 
formation in galaxies of various types and sizes, and to identify both 
commonalities and differences in their SFH across the LG to be able to make 
the comparison with cosmological surveys (Tolstoy 1998) more accurate.

The Local Group (LG) will provide a picture of the global star formation 
properties of galaxies with a wide variety of mass, metallicity, gas 
content etc., and will make a sample that ought to reflect the SFH of the 
Universe which can be compared to high redshift survys 
({\it e.g.}, Madau {\it et al.} 1998).  Initial comparisons suggest these 
different approaches do not yield the same results (Tolstoy 1998; Fukugita 
{\it et al.} 1998), but the errors are large due to the lack of {\it 
detailed} SFHs of nearby galaxies.  Careful determinations of the complete 
SFH of nearby galaxies is thus extremely important for understanding star 
formation properties of nearby galaxies in detail and also to make the 
connection to high-redshift studies.  It is still not clear if bursting 
star formation is common enough in LG dIs to account for a large population 
of compact emission line galaxies at intermediate redshifts.  If the 
conclusions 
of Madau {\it et al.} and the SFH of the LG still do not agree then 
it will be of critical importance for the field of high redshift galaxy 
research that this discrepancy is understood.  It might be that redshift 
surveys are strongly biased towards low mass dI galaxies undergoing bursts 
of star formation, and thus they are not accurate indicators of the 
dominant mode (by mass) of star formation in the Universe.

\section{Colour-Magnitude Diagram Analysis}

Stellar evolution theory provides a number of predictions, based on 
relatively well understood physics, of features expected in CMDs for 
different age and metallicity stellar populations (see Figure~1).  There 
are a number of clear indicators of varying {\it sfr}s at different times 
which can be combined to obtain a very accurate picture of the entire SFH 
of a galaxy.

\subsection{Main Sequence Turnoffs (MSTOs)}

If we can obtain deep enough exposures of the resolved stellar populations 
in nearby galaxies we can obtain the {\it unambiguous age information that 
comes from the luminosity of MSTOs}.  MSTOs can clearly distinguish between 
bursting star formation and quiescent star formation, ({\it e.g.} 
Hurley-Keller {\it et al.} 1998).  The age resolution that is possible does 
vary, becoming coarser going back in time.  Our ability to disentangle the 
variations in {\it sfr} from MSTOs also depends upon the the intensity of 
the past variations and how long ago they occured.  For ages less than 
about 1.5~Gyr it is possible to have detailed resolution on the 
10$-$100~Myr time scales.  As can be seen in Figure~1, the ages begin to 
crowd together for older populations.  Beyond about 8~Gyr ago the age 
resolution is on the scale of roughly a Gyr.  This of course means that a 
very short high intensity burst of star formation in this period will be 
indistinguishable from a lower {\it sfr} but longer lasting epoch of star 
formation.  As will be described in the following sections, however, there 
are other indicators in a CMD which help to narrow down the range of 
possible SFHs.

\hskip -1.5cm
\psfig{figure=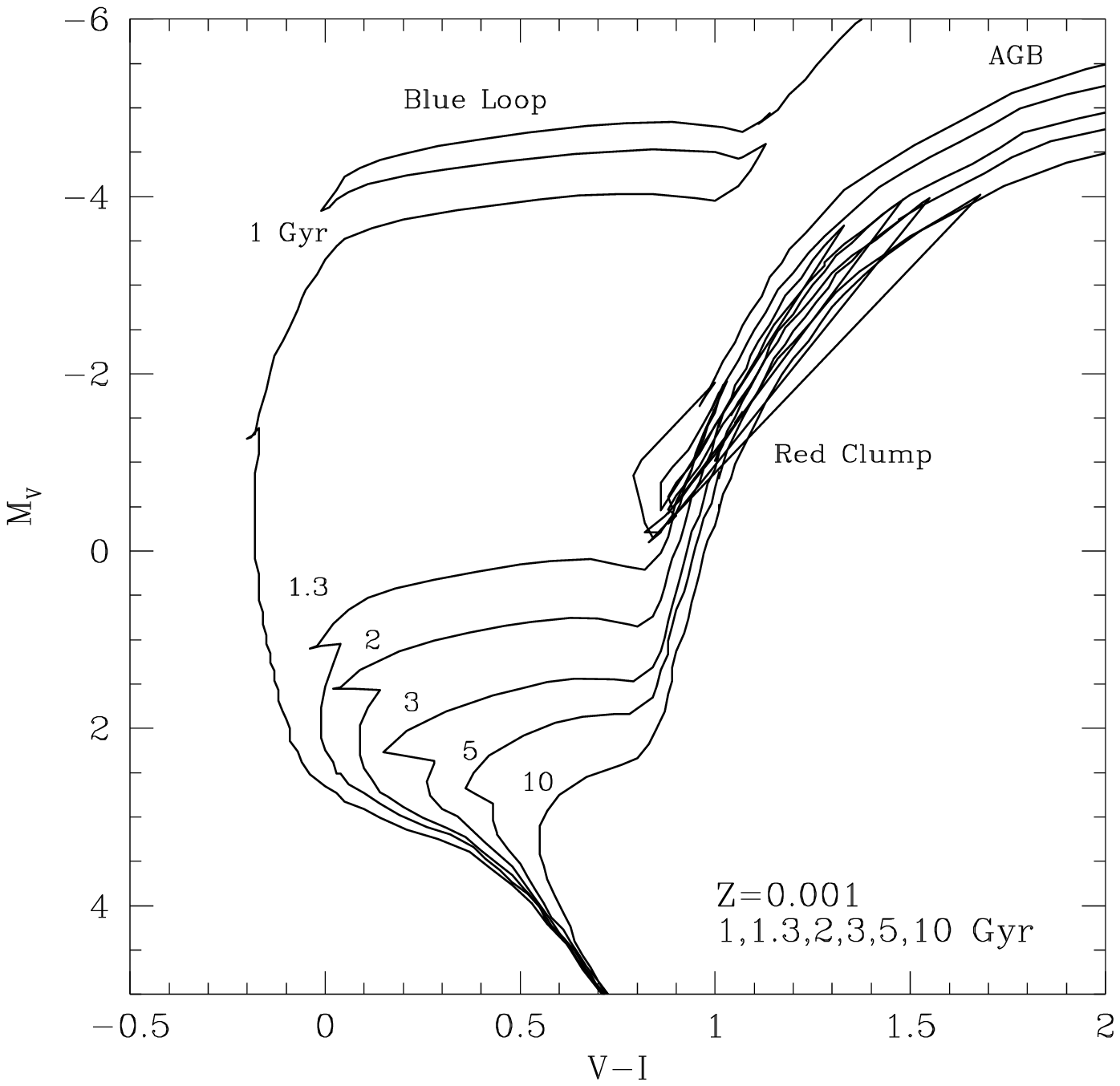,height=10cm,width=10cm}
\vskip -7.1cm
\hskip 7.cm
\begin{picture}(100,30)(0,0)
\put(0,0){\parbox{5.5cm}
{\small Figure~1: Isochrones for a single metallicity (Z=0.001) and a range 
of ages (from Bertelli {\it et al.} 1994), marked in Gyr, at the MSTOs.  
Isochrones were designed for single age globular cluster populations and 
are best avoided in the interpretation of composite populations, which can 
best be modeled using Monte-Carlo techniques ({\it e.g.} Tolstoy 1996).  }}
\end{picture}
\vskip4.cm

\subsection{The Core-Helium Burning Blue Loop Stars (BLs)}

Stars of certain metallicity and mass go on ``Blue Loop Excursions'', after 
they ignite He in their core (see Figure~1).  Stars in the BL phase can be 
several magnitudes brighter than when on the MS.  The luminosity of a BL 
star is fixed for a given age.  Subsequent generations of BL stars do not 
overlie each other as they do on the MS.  The shape and mass at which these 
``loops'' are seen in a CMD are a strong function of metallicity and age, 
in the sense that the lower the metallicity of the galaxy the further back 
in time an accurate SFH can easily be determined, which makes them 
especially useful for low metallicity systems such as dI galaxies.  ({\it e.g.}  
Dohm-Palmer {\it et al.} 1997a,b, 1998; Tolstoy et al. 1998).

\subsection{The Red Giant Branch (RGB)}

The RGB is a very bright evolved phase of stellar evolution, where the star 
is burning H in a shell around its He core.  For a given metallicity the 
RGB red and blue limits are given by the young and old limits 
(respectively) of the stars populating it (for ages $\simgt$1~Gyr).  As a 
stellar population ages the RGB moves to the red, and for constant 
metallicity, the blue edge is determined by the age of the oldest stars.  
However increasing the metallicity of a stellar population will also 
produce exactly the same effect as aging, and also makes the RGB redder.  
This is the age-metallicity degeneracy problem.  The result is that if 
there is metallicity evolution within a galaxy, it can be impossible to 
uniquely disentangle effects due to age and metallicity on the basis of the 
optical colours of the RGB alone, especially in low metallicity systems 
such as dI galaxies (cf.  Da Costa 1997)

\vskip.5cm
\hskip -.9cm
\psfig{figure=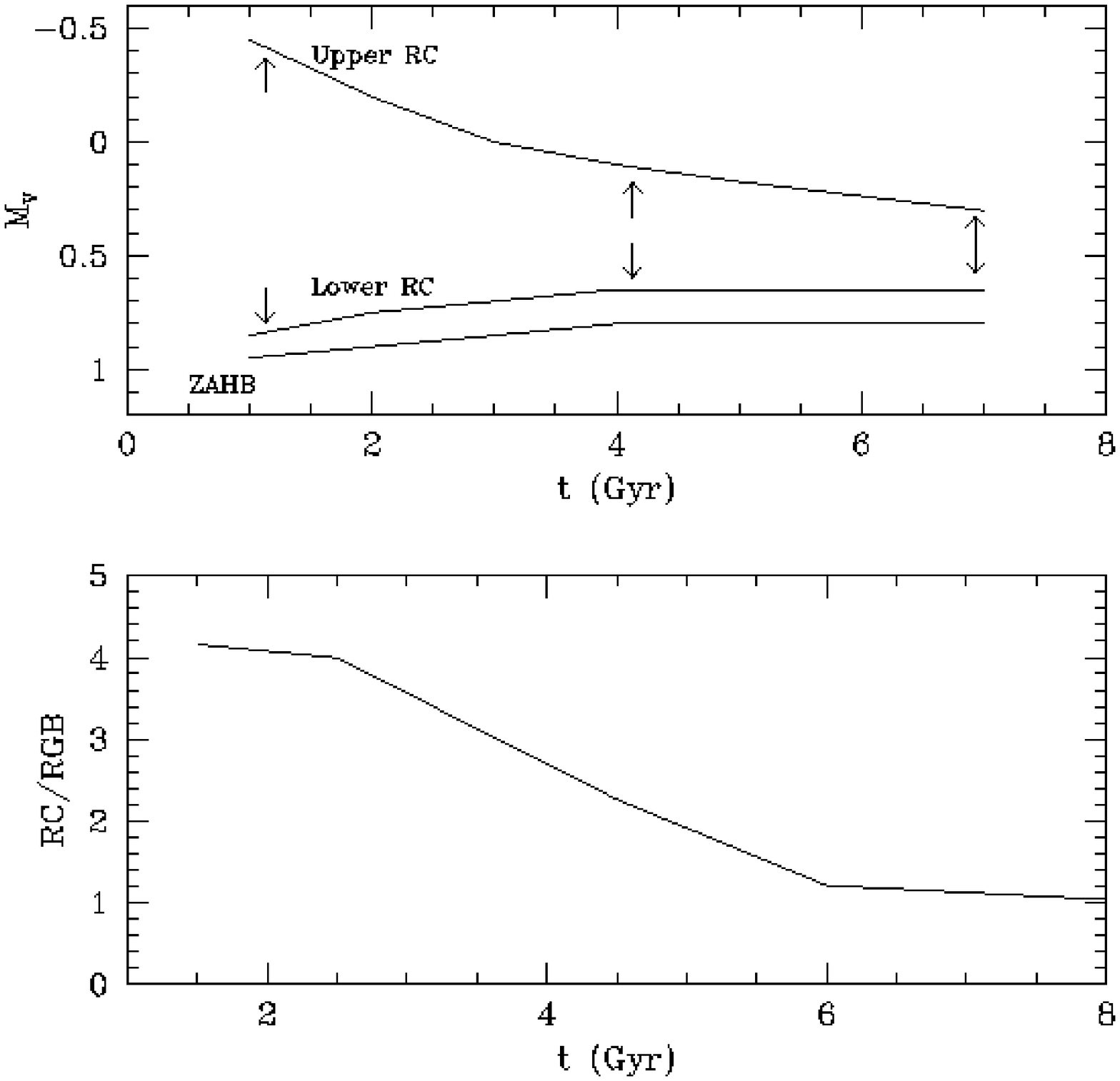,height=9.cm,width=8cm}
\vskip -6.8cm
\hskip 7.3cm
\begin{picture}(100,60)(0,0)
\put(0,0){\parbox{5.3cm}
{\small Figure~2: In the top panel is plotted the magnitude of the upper 
and lower edge of the RC versus age, in Gyr, from Caputo {\it et al.} 
(1995).  This shows the variation in the {\it extent} in M$_V$ of a RC with 
age, for a metallicity of Z=0.0004.  It can be clearly see that this extent 
is strong function of the age of the stellar population.  Also plotted is 
M$_V$ of the zero age HB against age.

In the bottom panel are plotted the results of running a series of 
Monte-Carlo simulations using stellar evolution models at Z=0.0004 (from, 
Fagotto {\it et al.}  1994) and counting the number of RC and RGB stars in the 
same part of the diagram, and thus we determine the expected ratio of 
RC/RGB stars versus age.  }}
\end{picture}
\vskip4.7cm

\subsection{The Red Clump/Horizontal Branch (RC/HB)}

Red Clump (RC) stars and their lower mass cousins, Horizontal Branch (HB) 
stars are core helium-burning stars, and their luminosity varies depending 
upon age, metallicity and mass loss ({\it e.g.}  Caputo {\it et al.}  1995).  The 
extent in luminosity of the RC can be used to estimate the age of the 
population that produced it, as shown in the upper panel of Figure~2.  This 
age measure is {\it independent of absolute magnitude and hence distance}, 
and indeed these properties can be used to determine an accurate distance 
measure on the basis of the RC ({\it e.g.}  Cole 1998).

The classical RC and RGB appear in a population at about the same time 
($\sim$ 0.9--1.5 Gyr, depending on model details), where the RGB are the 
progenitors of the RC stars.  The lifetime of a star on the RGB, t$_{RGB}$, 
is a decreasing function of M$_{star}$, but the lifetime in the 
RC, t$_{RC}$ is roughly constant.  Hence the ratio, t$_{RC}$ / t$_{RGB}$, 
is a decreasing function of the age of the dominant stellar population in a 
galaxy, and the ratio of the numbers of stars in the RC, and the HB to the 
number of RGB is sensitive to the SFH of the galaxy (Tolstoy {\it et al.}  1998; 
Han {\it et al.} 1997).  Thus, the higher the ratio, N(RC)/N(RGB), the 
younger the dominant stellar population in a galaxy, as shown in the lower 
panel of Figure~2.

The presence of a large HB population on the other hand (indicated by a 
high N(HB)/N(RGB) or even N(HB)/N(MS), is caused by a predominantly much 
older ($>$10~Gyr) stellar population in a galaxy.  The HB is the brightest 
indicator of very lowest mass (hence oldest) stellar populations in a 
galaxy.

\vskip -7.5cm
\hskip -1.cm
\psfig{figure=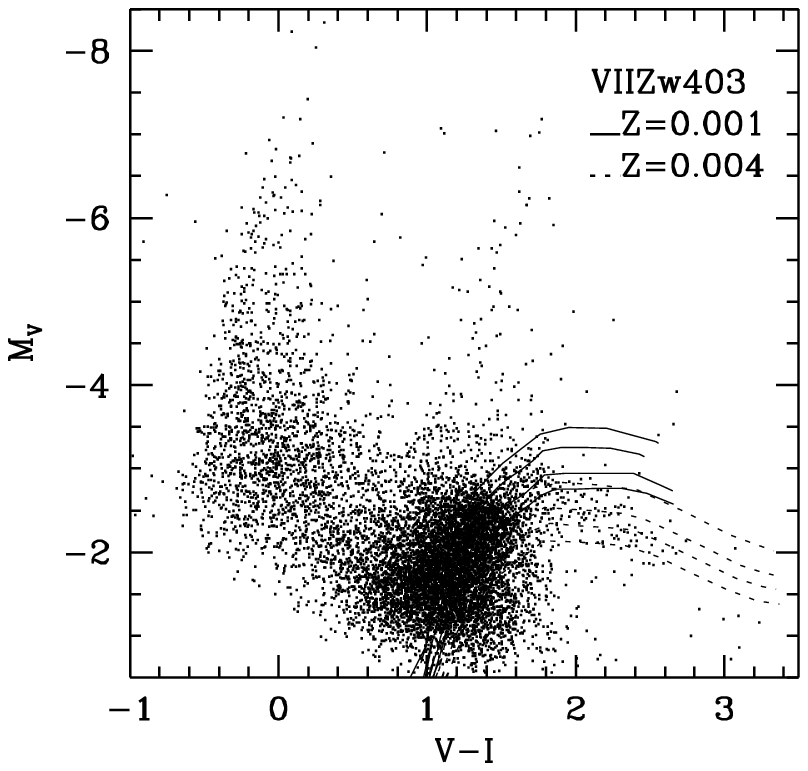,height=15cm,width=15cm}
\vskip -5.1cm
\hskip 6.8cm
\begin{picture}(100,30)(0,0)
\put(0,0){\parbox{5.8cm}
{\small Figure~3: EAGB isochrones \cite{b94} for metallicities, Z=0.001 and 
Z=0.004, are shown superposed on the observed CMD of VII~Zw403 \cite{L98}.  
For each metallicity the isochrones are for populations of ages 1.3, 2, 3, 
and 5 Gyrs, with the youngest isochrone being the brightest.  This shows 
the potential discriminant between the age and metallicity of older 
populations, if the models could be calibrated.}}
\end{picture}
\vskip 3.3cm

\subsection{The Extended Asymptotic Giant Branch (EAGB)}

The temperature and colour of the EAGB stars in a galaxy are determined by 
the age and metallicity of the population they represent.  However there 
remain a number of uncertainties in the comparison between the models and 
the data ({\it e.g.}  Gallart {\it et al.} 1994; Lynds {\it et al.} 1998).  It is 
very important that more work is done to calibrate these very bright 
indicators of past star formation events.  In Figure~3 theoretical EAGB 
isochrones (from Bertelli {\it et al.}  1994) are overlaid on the HST CMD of the 
post-starburst BCD galaxy, VII~Zw403, and we can see that a large 
population of EAGB stars is a bright indicator of a past high {\it sfr}, 
and the luminosity spread depends upon metallicity and the age of the {\it 
sfr}.

\subsection{Distance, Extinction \& Metallicity}

The accurate interpretation of the indicators described above depends 
critically upon having reliable estimates for the distance, the extinction 
and the metallicity of a galaxy.  Ideally we would also like to know if the 
extinction is patchy and on what scale, and what has been the evolution of 
the metallicity of the stellar population with time.  These three basic 
parameters, in conjunction with observational errors and incompleteness 
make the most significant impact on the properties of the CMD and hence the 
final SFH model (Tolstoy {\it et al.} 1998).  There are a number of 
difficulties in accurately determining these basic properties but they can 
be resolved with careful observation and analysis techniques.

\noindent{{\it The Distance}} is the most crucial parameter for accurate
analysis of a CMD, partly because it can easily be wrong by many orders of 
magnitude (for example the young, red supergiants can be mistaken for the 
RGB, if the observed CMD isn't deep enough to confirm the identification, 
{\it i.e.}  by detecting a RC or HB).  If the distance to a galaxy is incorrect 
this will result in the assumed ages of the different populations being 
completely wrong.  To be sure of the distance to small faint galaxies it is 
necessary to have a CMD which goes deep enough to extend below the 
unambiguous RC/HB strucutre (Tolstoy {\it et al.} 1998).

\noindent{{\it The Extinction}}, both internal to a galaxy and between us 
and a galaxy can affect the accurate analysis of a CMD.  If the extinction 
is incorrectly determined it will have the same effect as a distance error, 
and hence effect the reliability of the SFH models ({\it e.g.}  Gallagher {\it et 
al.} 1998).

\noindent{{\it Metallicity:}} When a galaxy makes stars, then the detritus 
of this process ({\it e.g.}, from SN explosions and stellar winds) make it 
unlikely that the galaxy can avoid metallicity evolution altogether.  
However, there is no concrete observational evidence that this is true, 
although abundance ratios of different elements do give us model dependent 
suggestions (Pagel 1994).  Determining the effect of metallicity evolution 
in a CMD is difficult.  It is impossible to determine a unique model based 
solely on the RGB because of age-metallicity degeneracy, and if metallicity 
evolution is neglected then the best model for that galaxy will typically 
be younger than if metallicity evolution were included ({\it e.g.}  Tolstoy {\it 
et al.} 1998).

\subsection{High Quality Imaging}

Typically the prime limiting factor in making an accurate analysis of a
CMD is image crowding, due to overlapping stellar images.
There is basically no way to ``correct'' for crowding
in a CMD, so all that can be done is to estimate the size of the
uncertainties that result from crowding, 
from incompleteness analysis ({\it e.g.} Tolstoy 1996),
and apply these to any results from the interpretation of the CMD.
Using $HST$ significantly reduces crowding problems, and in certain
cases, such as dwarf irregular galaxies, excellent seeing from the
ground will also be sufficient to overcome severe crowding problems.

\section{The Data: Recent HST Observations}

Thus the time is ripe for re-observing the the resolved stellar populations 
of all the galaxies in our LG.  Much of our detailed knowledge of the SFHs 
of galaxies beyond 1~Gyr ago comes from the Milky Way and its nearby dSph 
satellites or from $HST$ CMDs.  To date, the limiting factors have been 
crowding and resolution limits for accurate stellar photometry from the 
ground.  The large collecting area, the field field of view, and the 
extremely impressive image quality and stability of UT1 combined with the 
extremely good seeing attainable at Paranal makes it possible to obtain 
more accurate CMDs to fainter magnitudes, (ie.  do {\it better}) than 
$HST$.  These facilities provide unique opportunities to extend beyond our 
immediate vicinity and encompass the whole LG.  To date $HST$ has observed 
the resolved stellar populations in variety of nearby galaxies ({\it e.g.}, dE: 
NGC~147, Han {\it et al} 1997; Irr: LMC, Geha {\it et al.} 1998; Spiral: 
M~31, Holland {\it et al.} 1996; M~33, Sarajedini {\it et al.}  1998; BCD: 
VII~Zw~403, Lynds {\it et al.} 1998; NGC~1569, Greggio {\it et al.} 1998; 
dI: Leo~A Tolstoy {\it et al.} 1998; dSph: Leo~I, Gallart {\it et al.} 
1999).  Every galaxy which has been looked at carefully something new has 
been learnt.

An $HST$ program was initiated by S
killman ({\it e.g.}  Skillman {\it et al.}  1998), 
observing a sample of four nearby dI galaxies (Sextans~A, Pegasus, Leo~A \& 
GR~8), using four orbits of telescope time per galaxy, in three filters 
(effectively B, V and I).  The results have been dramatic and illustrate 
the tremendous advances possible, even with short exposures, when crowding 
has been virtually illuminated.  Here I am going to provide a summary of 
this program.

\hskip -0.5cm \centerline{
\psfig{figure=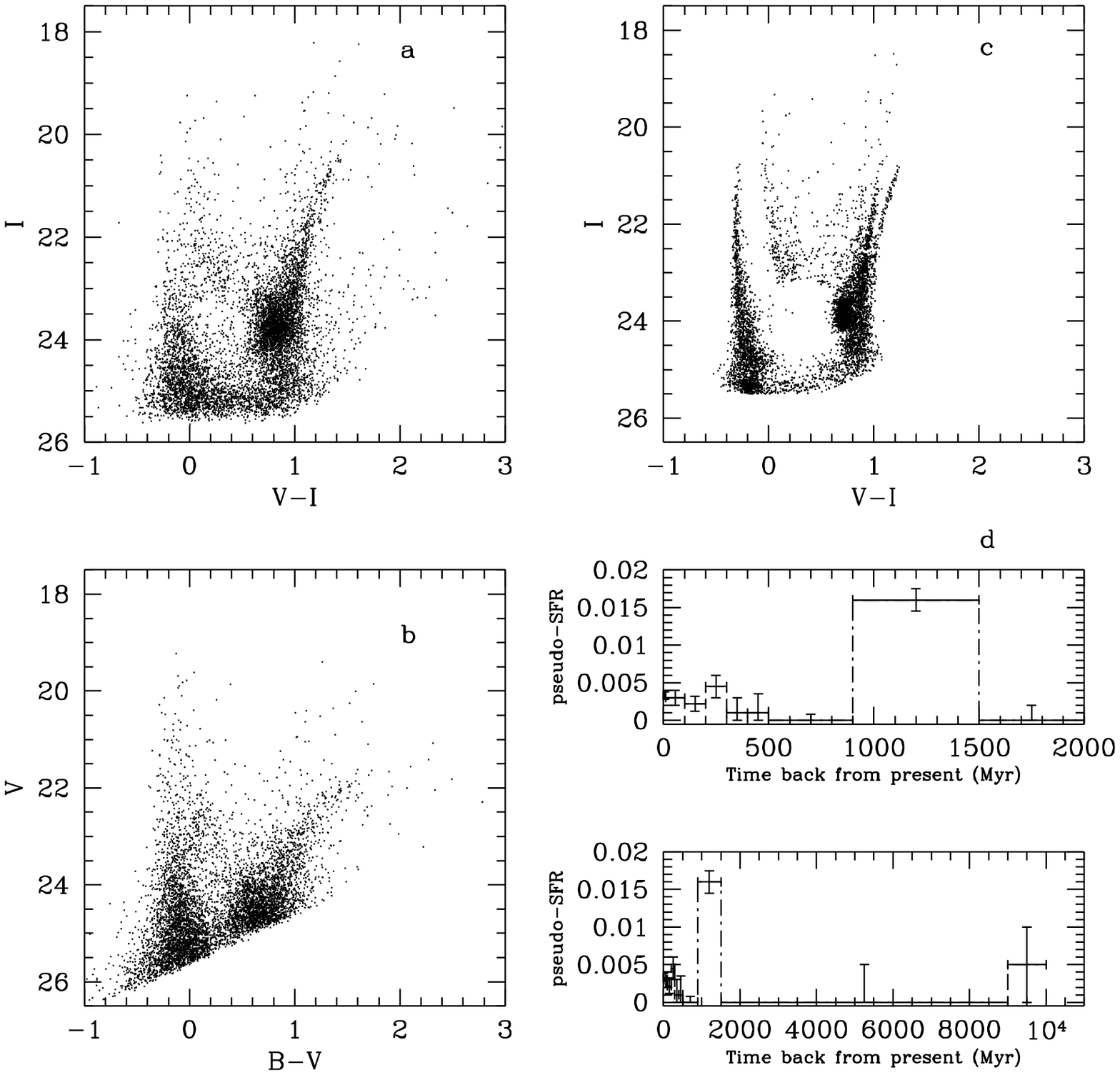,height=15cm,width=15cm}} \vskip0.3cm
\begin{picture}(300,30)(0,0)
\put(0,0){\parbox{13cm}
{\small Figure~4: Here we show the results for the analysis of the 
HST/WFPC2 data of Leo~A (Tolstoy {\it et al.}  1998).  In a.  is the V$-$I, I 
CMD, 1 orbit exposure time per filter.  In b.  is the B$-$V, V CMD, 2 
orbits in B.  In c.  is the best match Monte-Carlo simulation model (in 
V$-$I, I) found for these data convolved with the theoretical measurement 
error distribution, and in d.  is the SFH that created the model CMD which 
best matches these data. See Tolstoy {\it et al.} 1998 for more details. }}
\end{picture}
\vskip 1.5cm

\subsection{Leo~A}

Leo~A ($\equiv$ DDO~69, Leo III, 
UGC~5364) is a nearby gas-rich Magellanic dI galaxy.
Analysis of the 
WFPC2 CMDs of the resolved stellar population (see Figure~4) 
resulted a new distance and an accurate SFH to be determined for this 
extremely metal-poor galaxy (Tolstoy {\it et al.} 1998).  From the 
position of the RC, BLs and the tip of the RGB, a distance modulus, 
m$-$M=24.2$\pm$0.2, or 690 $\pm$ 60 kpc, was obtained which places Leo~A 
firmly within the LG.  The interpretation of the WFPC2 CMDs at this new 
distance based upon extremely low metallicity (Z=0.0004) theoretical 
stellar evolution models suggests that this galaxy is predominantly young, 
{\it i.e.} $<$~2~Gyr old.  A major episode of star formation 900$-$1500~Gyr 
ago can explain the RC luminosity and also fits in with the interpretation 
of the number of anomalous Cepheid variable stars seen in this galaxy.  The 
presence of an older, underlying globular cluster age stellar population 
could not be ruled out with these data, however, using the currently 
available stellar evolution models, it would appear that such an older 
population is limited to no more than 10\% of the total star formation to 
have occured in the centre of this galaxy. Using theoretical 
models of the chemical evolution
of dwarf galaxies (Ferrara \& Tolstoy 1999) however, it is clear that
even though this galaxy is extremely metal poor in order for it to build
up the current metallicity it must 
contain a significant 
underlying older stellar population, perhaps in an
extended outer halo of older stars. Of course neither the chemical
evolution models nor the existing CMDs can distinguish between an old
population which formed in a large burst, or more sedate and
roughly constant rate through-out a longer time.

\subsection{Pegasus}

The Pegasus dI galaxy is a gas-rich system in the LG, it is possibly
an example of a transition object between 
a dwarf galaxy dominated by current
star-formation and one dominated by past star formation.
Analysis of the 
WFPC2 CMD of the Pegasus dI (see Figure~5) reveals a young main 
sequence (MS), a well populated RGB, a small number of EAGB stars and near 
the faint limits of the data a populous RC (Gallagher {\it et al.} 1998).  
The young stellar component is clustered in two centrally-located clumps, 
while older stars form a more extended disk or halo.  The colours of the MS 
require a relatively large extinction (A$_V =$ 0.47 mag), and the 
(extinction corrected) mean colour of the well-populated RGB is relatively 
blue, consistent with a moderate metallicity young, or older and more 
metal-poor stellar population.  The distance of Pegasus was revised to be 
760 kpc, taking account of the higher reddening.  The RGB has significant 
width in colour, implying a range of stellar ages and/or metallicities.  A 
SFH which was consistent with the data is one in which the {\it sfr} was 
higher, by a factor of 3-4, about 1~Gyr ago.  It was impossible to 
constrain the SFH beyond 1~Gyr ago, as seen by the large error bars in 
Figure~5d, without better information on stellar metallicities and deeper 
photometry.  The youngest model consistent with the data contains stars 
with constant metallicity of Z$=$0.001 which mainly formed 2-4 Gyr ago.  If 
stellar metallicity declines with increasing stellar age, then the older 
ages extend up to $\sim$8~Gyr.  However, even at its peak of star 
forming activity, the intermediate-age dominated model for the Pegasus 
dwarf most likely remained relatively dim with M$_V \approx -$14.

\hskip -0.5cm \centerline{
\psfig{figure=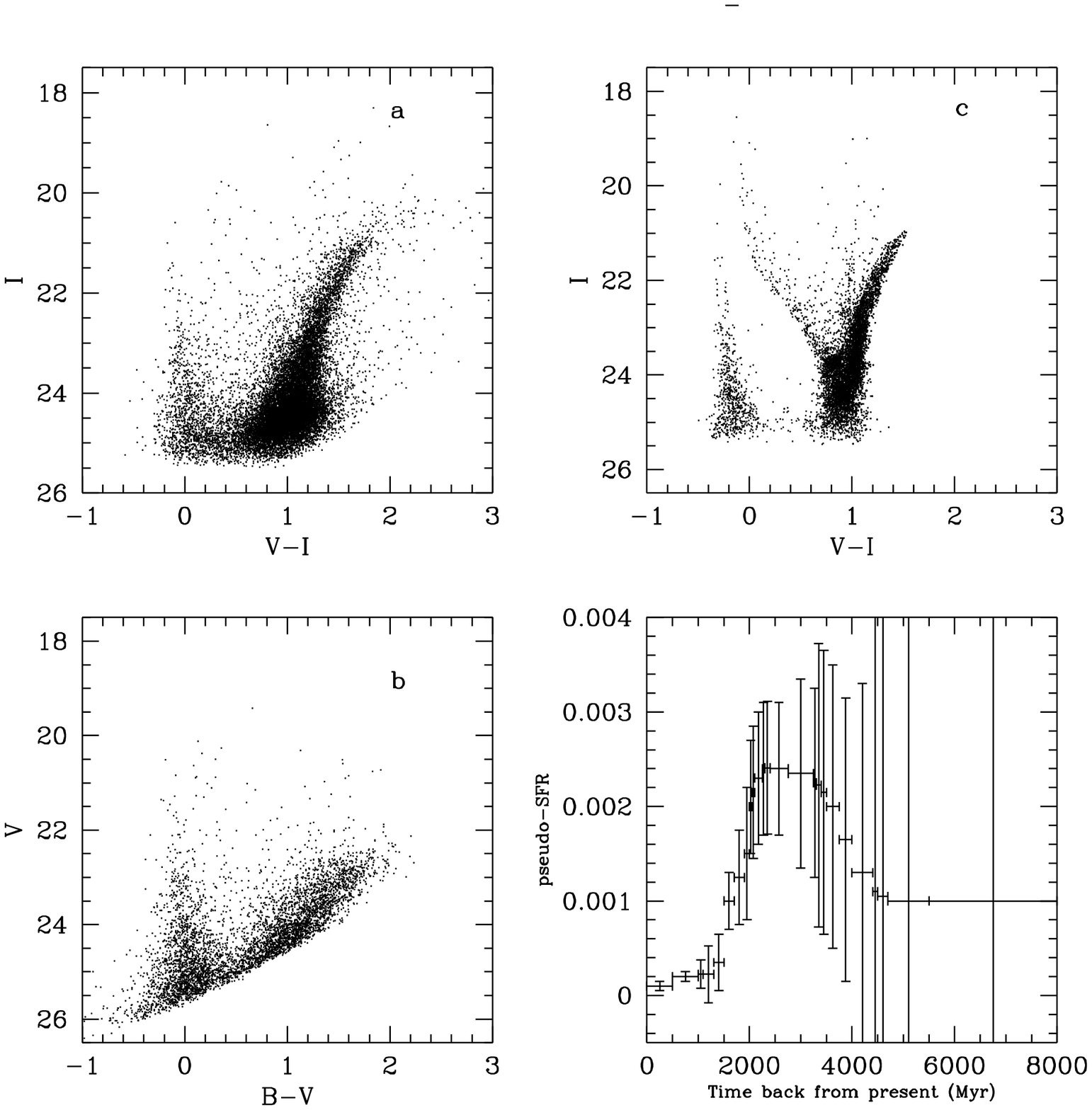,height=15cm,width=15cm}} \vskip0.3cm
\begin{picture}(300,30)(0,0)
\put(0,0){\parbox{13cm}
{\small Figure~5: Here we show the results for the HST/WFPC2 analysis of 
Pegasus (Gallagher {\it et al.}  1998).  In a.  is the V$-$I, I CMD, 1 orbit 
exposure time per filter.  In b.  is the B$-$V, V CMD, 2 orbits in B.  In 
c.  is the best match Monte-Carlo simulation model, in V$-$I, I, found for 
these data ({\it excluding the RC}) and convolved with the theoretical 
measurement error distribution, and in d.  is the SFH that created the 
model CMD which best matches the data. See Gallagher {\it et al.} 1998 for more
details.}}
\end{picture}
\vskip 1.5cm

\subsection{Sextans~A}

Sextans A (DDO 75, A 1008-04) is a gas-rich dI galaxy with
active star formation, which is located on the periphery of the Local
Group at a distance of 1.44 Mpc.
The CMD of Sextans~A shows several clearly separated populations that align 
well with stellar evolution model predictions for a low metallicity system 
(Dohm-Palmer {\it et al.} 1997b).  The SFH from the MS+BL stars was 
determined over the last 600~Myr.  This is the first time a BL sequence has 
been clearly identified in a CMD.  The distribution of the BL stars was 
combined with {\it sfr} calculations to determine the spatial variation of 
the star formation across Sextans~A with time.  The modelling concludes 
that in the past 50 Myr, Sextans A has had an average {\it sfr} that is 
$\sim10$ times that of the average {\it sfr} over the history of the 
galaxy.  This current activity is highly concentrated in a young region in 
the Southeast roughly 25 pc across.  This coincides with the brightest HII 
regions and the highest column density of HI.  This one region contains 
half of all the current star formation activity within the field of view.  
Between the ages of 100 and 600 Myr ago, the star formation has been 
roughly constant at slightly above the average value.  There are regions 
(200$-$300pc across) with a factor of $\sim 5$ enhancement in {\it sfr} 
with a duration of 100$-$200 Myr.

\hskip -0.5cm \centerline{
\psfig{figure=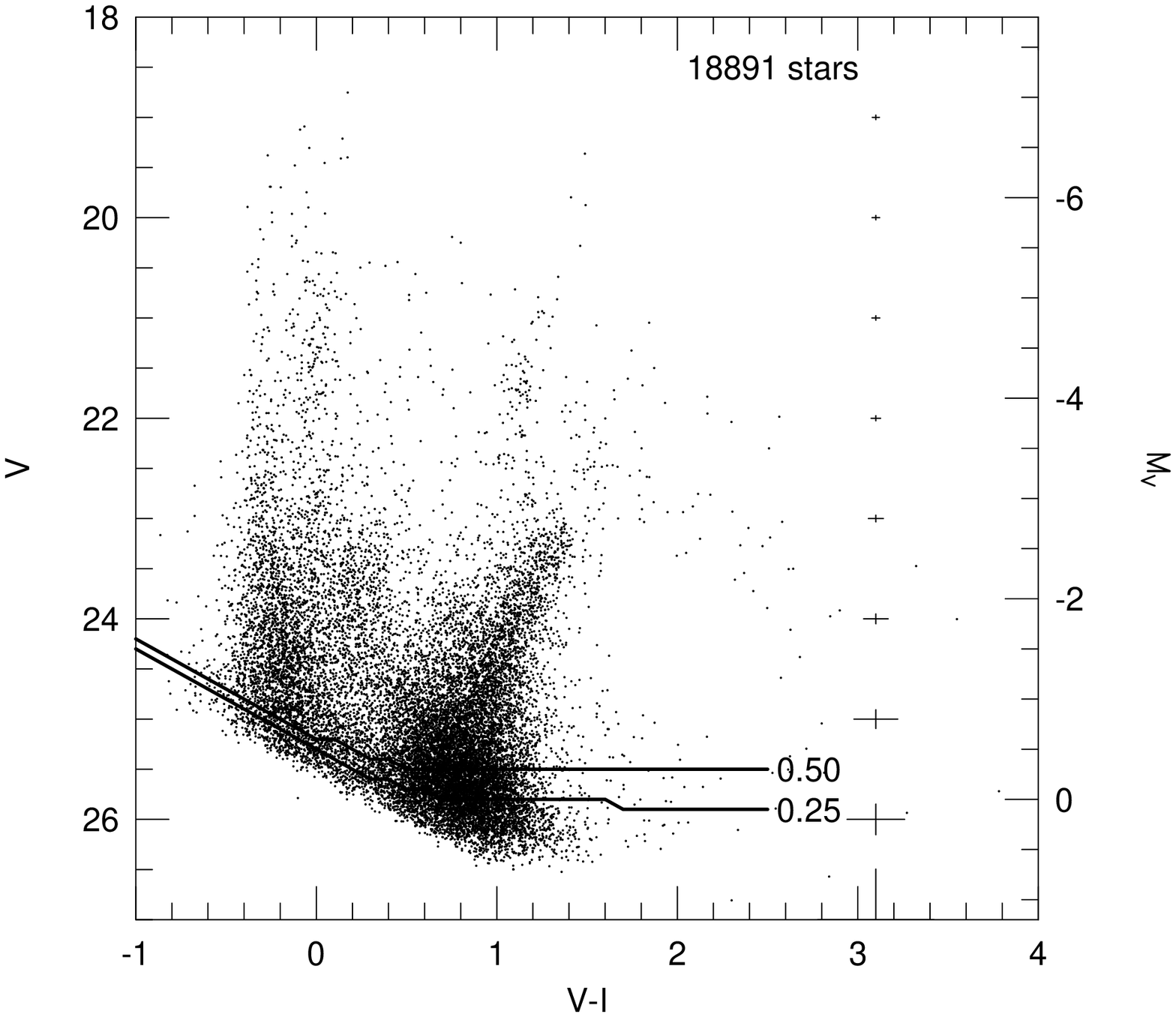,height=7.5cm,width=7.5cm}
\psfig{figure=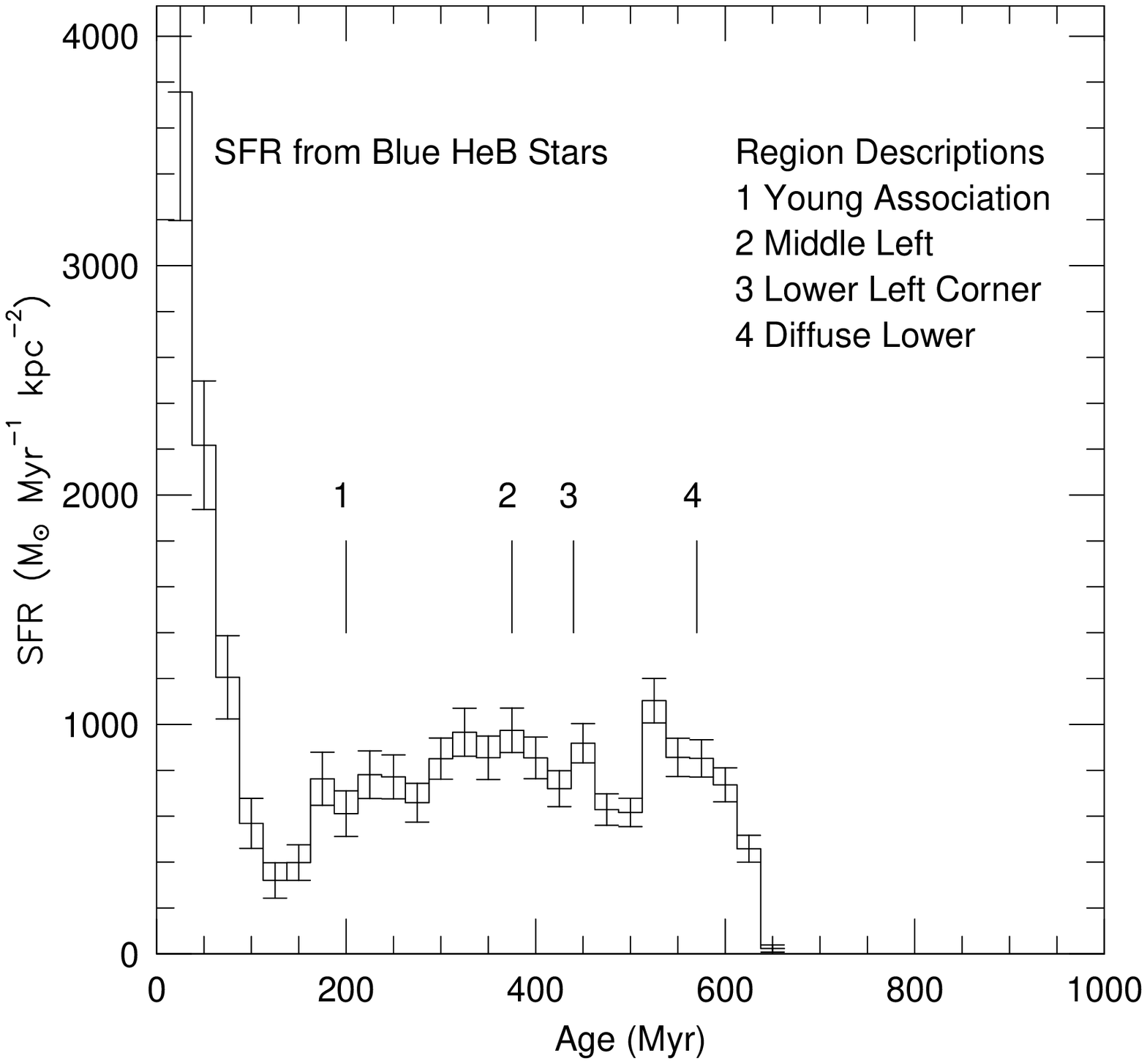,height=7.5cm,width=7.5cm}} \vskip-1.3cm
\begin{picture}(300,30)(0,0)
\put(0,0){\parbox{13cm}
{\small Figure~6: Here we show the results for the HST/WFPC2 analysis of 
Sextans A (Dohm-Palmer {\it et al.}  1997b).  On the left is the V$-$I, I CMD, 1 
orbit exposure time per filter. On the right is the SFH from the MS+BL
which best matches the data. See Dohm-Palmer {\it et al.} for more details.}}
\end{picture}
\vskip .8cm

\subsection{GR~8}

The final galaxy looked at with WFPC2 in this sample is 
GR 8 (DDO 155, UGC 8091) is a gas-rich dI galaxy with active star
formation, on the edge of the LG, at a distance of 2.2~Mpc
(Dohm-Palmer {\it et al.}  1998).
It is slightly more compact and distant than 
the other galaxies in the Skillman sample, and so crowding was more of a 
problem in the analysis of the WFPC2 data.
Artificial star tests showed the data to be 50\% complete to 
$V=26.3$, $B=25.4$, and $I=25.2$.  The CMD shows a well defined population 
with a very young MS ($< 10$ Myr), and an RGB as old as several Gyr.  A 
distance estimate, based on the tip of the RGB, was found to be in 
excellent agreement with the Cepheid determination of $\mu = 26.75 \pm 
0.35$, or 2.2 Mpc (Tolstoy {\it et al.} 1995).  There is also evidence for 
an extended stellar ``halo'' beyond the HI distribution.  Based on the MS 
and BL luminosity function the {\it sfr} over the past 500 Myr was 
estimated to have been fairly constant, with up to 60\% variations.  The 
star formation appears to occur in super-association size regions (100-200 
pc across), which last $\sim 100$ Myr.  These regions come and go with no 
obvious pattern, except that they seem to concentrate in the current 
locations of HI clumps.  This suggests that the HI clumps are long lived 
features that support several star forming events over time.  

\hskip -0.5cm \centerline{
\psfig{figure=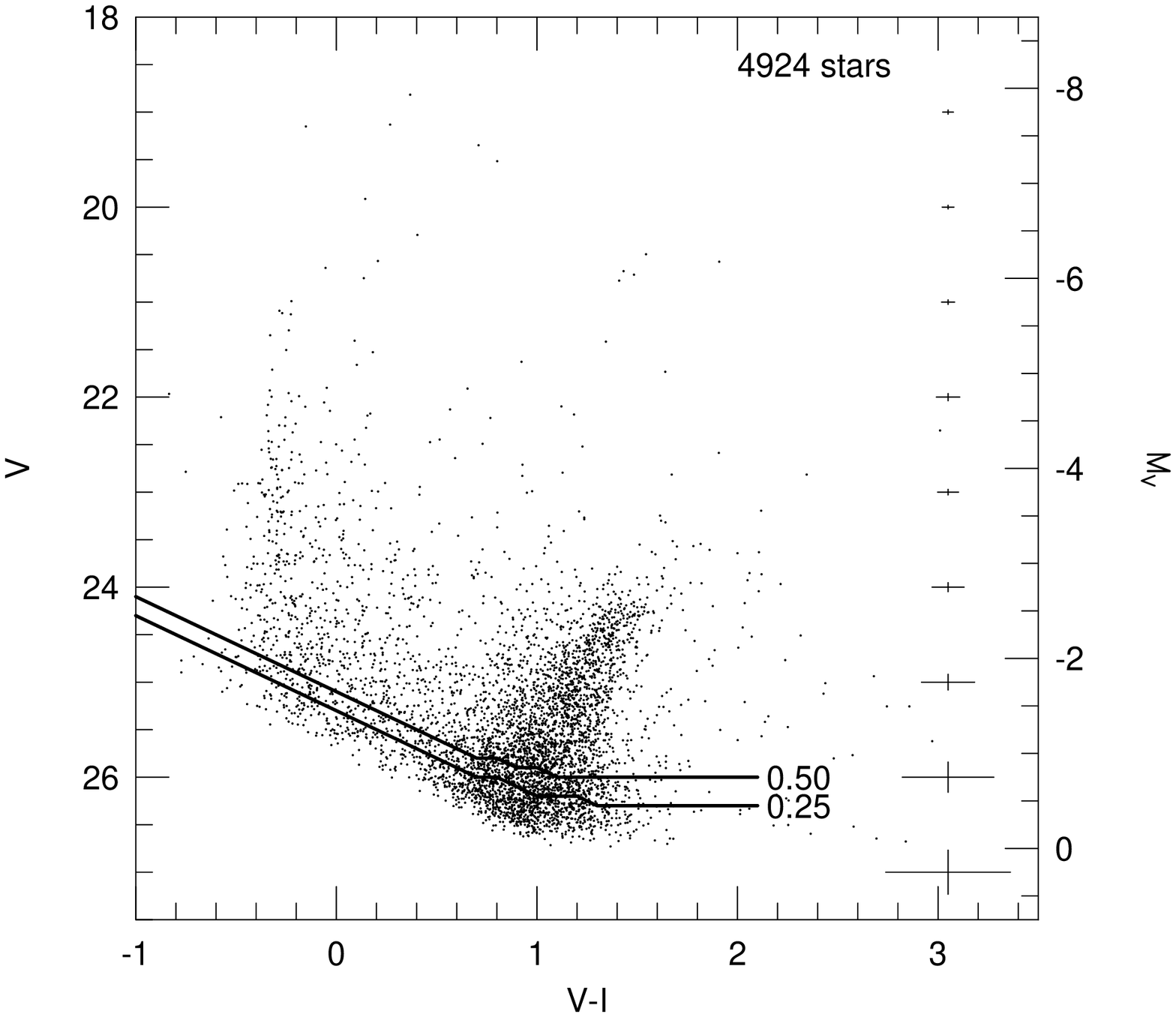,height=7.5cm,width=7.5cm}
\psfig{figure=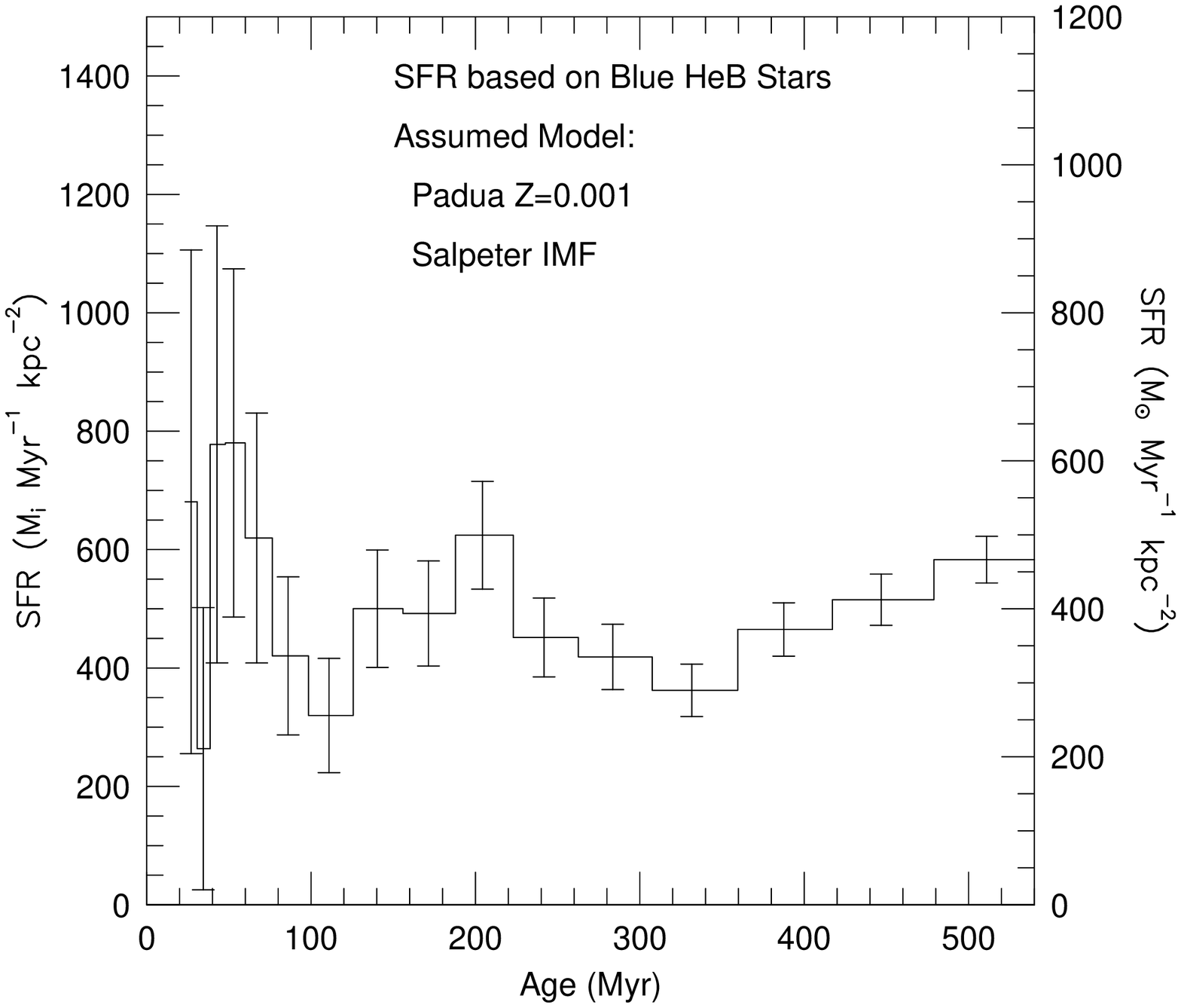,height=7.5cm,width=7.5cm}} \vskip-1.3cm
\begin{picture}(300,30)(0,0)
\put(0,0){\parbox{13cm}
{\small Figure~7: Here we show the results for the HST/WFPC2 analysis of 
GR~8 (Dohm-Palmer {\it et al.}  1998).  On the left is the V$-$I, I CMD, 1 orbit 
exposure time per filter.  On the right is the SFH from the MS+BL
which best matches the data. See Dohm-Palmer {\it et al.} for more details. }}
\end{picture}
\vskip .5cm

\section{The Scenario}

There appear to be 
a diversity of star formation properties between apparently 
similar galaxy types. This appears to correlate with neither the 
metallicity of a galaxy nor its gas content. In this sample there is no 
direct evidence (MSTOs) for any discrete strong bursts of star 
formation. 
Comparing the results of the four dI galaxies in the HST 
sample of Skillman: Sextans A, Pegasus DIG, Leo A, \& GR~8 
a trend of higher 
{\it sfr} per area with larger $M_{HI}/L_B$ was 
found (Dohm-Palmer {\it et al.} 1998).  
Although Pegasus and Leo~A have clearly had higher 
global star formation rates in the past, whether this took place over 
several Gyr or in a short discrete burst cannot be discerned with these 
data. 
Leo A is the best nearby candidate for a primeaval galaxy. If 
there was an epoch during which galaxies had frequent bursts it was 
beyond z=0.1. Deeper observations of more galaxies
will push this limit back in 
redshift, and make the connection to the faint blue galaxy population
more
convincing.
The giant galaxies (M~31, the Galaxy, M~33) have large old populations, and
must contain the bulk of star formation in the LG since 
they contain something like 90\% of the stellar mass, 
but this does not mean that they have always dominated in luminosity 
as completely as they do now. This is why it is important 
to look for detailed
evidence of bursting behaviour in the most common (by number) galaxies
in the LG - the late type irregular galaxies - 
by looking at the fossil record of the old/intermediate age  
resolved stellar populations.

\acknowledgments

I gratefully acknowledge the organisers for a generous grant, and I
also thank my collaborators on the ``HST dwarf galaxy 
consortium'', namely, Evan Skillman, Robbie Dohm-Palmer, Jay Gallagher, 
Andrew Cole, Mario Mateo, Abi Saha \& John Hoessel 
who have been involved 
in all of the projects I have reported on here.

\end{document}